\documentstyle[emulateapj]{article}

\begin{document}

\title{{\it FUSE} Observations of Molecular Hydrogen in Translucent
Interstellar Clouds: The Line of Sight Toward HD 73882}


\author{T. P. Snow\altaffilmark{1},
B. L. Rachford\altaffilmark{1},
J. Tumlinson\altaffilmark{1},
J. M. Shull\altaffilmark{1},
D. E. Welty\altaffilmark{2},
W. P. Blair\altaffilmark{3},
R. Ferlet\altaffilmark{4},
S. D. Friedman\altaffilmark{3},
C. Gry\altaffilmark{5,6},
E. B. Jenkins\altaffilmark{7},
A. Lecavelier\altaffilmark{4},
M. Lemoine\altaffilmark{8},
D. C. Morton\altaffilmark{9},
B. D. Savage\altaffilmark{10},
K. R. Sembach\altaffilmark{3},
A. Vidal-Madjar\altaffilmark{4},
D. G. York\altaffilmark{2},
B.-G. Andersson\altaffilmark{3},
P. D. Feldman\altaffilmark{3},
and H. W. Moos\altaffilmark{3}
}

\altaffiltext{1}{CASA, University of Colorado, Campus Box 389, Boulder, CO
80309}
\altaffiltext{2}{Astronomy and Astrophysics Center, University of Chicago,
5640 S. Ellis Ave., Chicago, IL 60637}
\altaffiltext{3}{Department of Physics and Astronomy, The Johns Hopkins
University, 3400 N. Charles St., Baltimore, MD 21218}
\altaffiltext{4}{Institut d'Astrophysique de Paris, CNRS, 98bis, Blvd.
Arago, Paris, F-75014, France}
\altaffiltext{5}{Laboratoire d'Astronomie Spatiale, B.P. 8, 13376 Marseille
cedex 12, France}
\altaffiltext{6}{ISO Data Center, ESA Astrophysics Division, PO Box 28080
Madrid, Spain}
\altaffiltext{7}{Princeton University Observatory, Princeton, NJ 08544}
\altaffiltext{8}{DARC, UMR 8629 CNRS, Observatorie de Paris, F-92195 Meudon,
France}
\altaffiltext{9}{Herzberg Institute of Astrophysics, National Research
Council, 5071 West Saanich Road, Victoria, BC V8X 4M6, Canada}
\altaffiltext{10}{Department of Astronomy, University of Wisconsin, 475 N.
Charter St., Madison, WI 53706}

\begin{abstract}
We report the results of initial {\it FUSE} observations of
molecular hydrogen (H$_2$) in translucent clouds. These clouds
have greater optical depth than any of the diffuse clouds previously
observed for far-UV H$_2$ absorption, and provide new insights
into the physics and chemistry of such regions.  Our initial
results involve observations of HD 73882, a well-studied
southern hemisphere star lying behind substantial interstellar
material ($E_{B-V}$ = 0.72; $A_V$ = 2.44).  We find a total
H$_2$ column density, N(H$_2$) = 1.2 $\times$ 10$^{21}$ cm$^{-2}$,
about three times larger than the values for diffuse clouds previously
measured in the far-UV.  The gas kinetic temperature indicated by the
ratio N($J$=1)/N($J$=0) is 58 $\pm$ 10 K.  With the aid of
ground-based data to calculate an appropriate multi-component curve
of growth, we have determined column densities for all rotational
levels up to $J$ = 7.  The $J$ $\geq$ 2 states can be reasonably
fitted with a rotational excitation temperature of 307 $\pm$ 23 K.
Both the kinetic and rotational temperatures are similar to those
found in previous investigations of diffuse clouds.  The ratios of
carbonaceous molecules to hydrogen molecules are also similar to
ratios in diffuse clouds, suggesting a similar chemistry for this
line of sight.
\end{abstract}


\section{Introduction}

Molecular hydrogen is the most abundant molecular species in the Galactic
interstellar medium.  While H$_2$ is an important constituent of diffuse
interstellar clouds with visual extinctions $A_V$ less than about
1 magnitude, it becomes the dominant form of hydrogen in the so-called
``translucent'' clouds, which are characterized by $A_V$ in the range
1--5 mag.  H$_2$ dominates the dynamics, chemistry, and physics of denser
clouds, and is the basic raw material for star formation.

Due to the homonuclear structure of H$_2$ and its lack of a dipole
moment, ro-vibrational transitions within the electronic ground
state are quadrupolar with low spontaneous emission coefficients.
They are therefore very difficult to observe except where special
excitation conditions produce detectable ro-vibrational emission in
the near-IR (e.g. Gautier et al.\ 1976) or from a space-borne
observatory such as ISO, when a long enough line of sight ($A_V$
$\approx$ 20) allows detection of the mid-IR pure
rotational emission from warm diffuse gas (Falgarone et al.\ 2000).
Cold H$_2$, the dominant form, can be widely observed through
electronic transitions in the far UV ($\lambda$ $<$ 1150 \AA ),
or with difficulty in the IR, where the weak quadrupole transition
can be observed in absorption only for very high column densities
(Lacy et al. 1994).

Previous instruments capable of resolving the far-UV absorption
bands of H$_2$ include the {\it Copernicus} mission
(Savage et al.\ 1977; Spitzer \& Jenkins 1975), the {\it IMAPS}
experiment (Jenkins \& Peimbert 1997), and the {\it ORFEUS}
spectrometer (Richter et al.\ 1998; de Boer et al.\ 1998).
All of those, however, were limited to relatively bright stars
with $A_V$ $\la$ 1.

The {\it FUSE} observatory (Moos et al.\ 2000) is well suited
for observations of cold H$_2$ in the diffuse and translucent
interstellar medium, owing to its high throughput in the wavelength
region from 905 \AA\ to 1187 \AA, encompassing the strong Werner and
Lyman bands of H$_2$.  A survey of H$_2$ in these regions was therefore
identified by the {\it FUSE} PI team as a project of high priority.

We have assembled a list of 35 lines
of sight, which sample a wide variety of interstellar environments.
These lines of sight, with 0.3 mag $\la$ $E(B-V)$ $\la$ 1.1 mag, are also
characterized by a range in extinction properties (determined both from
$R_V$ and from the shape of the far-UV extinction curve).  Most of the
lines of sight have information on molecular abundances from previous optical
and/or mm-wave data.

In conjunction with the {\it FUSE} observing program, we and several
collaborators are obtaining additional data
%for these lines of sight
using various ground-based telescopes.  We have obtained very
high-resolution (R $\sim$ 150,000--250,000) spectra of interstellar
\ion{K}{1}, \ion{Na}{1}, \ion{Ca}{2}, and CH absorption along most of
the lines of sight to understand the cloud velocity structure (Welty,
Morton, \& Snow 2000).  We (led by DGY) are obtaining moderately
high-resolution (R $\sim$ 50,000), very high S/N, nearly complete
optical spectra to measure many of the diffuse interstellar bands, and
the molecules CH, CH$^+$, CN, C$_2$, and C$_3$.  T. Oka and B. J. McCall
are pursuing near-IR measurements of H$_3$$^+$, and we and F. Chaffee
are conducting near-IR observations to compare
grain mantle features such as water ice, the 3.4-$\mu$m hydrocarbon band,
and the silicate feature at 9.6 $\mu$m, with the UV data on dust
extinction and gas-phase depletions.
The ultimate goals of this program are to fully understand the masses of
dense interstellar clouds, to probe the chemistry and physics of translucent
clouds, and to probe the transition region between diffuse and dense
interstellar clouds.

In this {\it Letter}, we describe the first observations of H$_2$
in a translucent cloud line of sight, toward the star HD 73882.
The following sections describe the properties of the line of sight deduced
from previous observations (\S 2), the {\it FUSE} observations and our
analysis of the data (\S 3), and the insights gained from these new
data (\S 4).

\section{The Line of Sight toward HD 73882}

The star HD 73882 is well known in the translucent cloud community
because it is of early spectral type (O8.5V), it is bright enough
($V$ = 7.27) to allow high-S/N observations at optical wavelengths,
and it has sufficient foreground dust ($E_{B-V}$ = 0.72, $A_V$ = 2.44)
and gas to have large column densities of many atomic and molecular
species.  Fitzpatrick \& Massa (1986, 1988, 1990) derived the UV
extinction curve and the column density of atomic hydrogen from
low-resolution {\it IUE} spectra.  The extinction curve shows a steep
far-UV rise with significant curvature, generally similar to other
``dense cloud'' curves, as characterized by Cardelli, Clayton, \&
Mathis (1988) -- a conclusion reached previously for this star by
Massa, Savage, \& Fitzpatrick (1983).  However, both the ratio of
total to selective extinction ($R_V$ = 3.9) and the wavelength of
maximum polarization (Serkowski, Mathewson, \& Ford 1975) suggest
that small grains are not as important in this case as in most other
molecular-cloud lines of sight.  The molecular column densities listed
in Table 1 support the view that this line of sight is dominated by
one or more dense clouds, consistent with translucent cloud models
(e.g. van Dishoeck \& Black 1986).

A series of papers have provided ground-based data on CO, CH, CN,
CH$^{\rm +}$, and C$_2$ (van Dishoeck et al.\ 1991; Gredel et
al.\ 1993; see Table 1).  The CO emission line profiles reveal three
distinct components toward HD 73882, at LSR velocities of +5.9, +8.9,
and +11.2 km s$^{-1}$.  The absorption from the other molecular species,
however, is seen only in a single component near +4.5 km s$^{-1}$.
Using the 3.6-m reflector
and coud\'{e} echelle spectrometer at ESO and the 0.9-m coud\'{e} feed
telescope and coud\'{e} spectrograph at KPNO, we have obtained very
high-resolution ($R$ $\sim$ 150,000--250,000) spectra of \ion{Na}{1},
\ion{K}{1}, and \ion{Ca}{2} absorption toward HD 73882.  A
profile-fitting analysis of the \ion{K}{1} and \ion{Na}{1} lines reveals
no fewer than 20 velocity components, spread over 55 km s$^{-1}$.
However, the absorption is dominated by three narrow ($b$ $\sim$ 0.7 km
s$^{-1}$), blended components at $v_{\rm LSR}$ $\approx$ +2.7, +4.3, and
+5.9 km s$^{-1}$, with significant contributions from two others at +7.7
and +11.0 km s$^{-1}$.  The blend centered at +4.3 km s$^{-1}$ probably
corresponds to the single (unresolved at $R$ $\sim$ 60,000---100,000)
absorption component seen in CH, CN, CH$^+$, and C$_2$, and to the CO
emission component at +5.9 km s$^{-1}$.  The two slightly weaker
atomic components may correspond to the two other CO
components at +8.9 and +11.2 km s$^{-1}$.  (Note that the CO column
density quoted in Table 1 is only for the +5.9 km s$^{-1}$ component.)
The numerous weaker features seen in the optical data may
arise in outlying less dense clouds.

\section{{\it FUSE} Observations and Data Analysis}

Our {\it FUSE} spectrum of HD 73882 is derived from time-tagged
observations over the course of 8 orbits on 1999 Oct 30.  Several
``burst'' events occurred during the observation (Sahnow et al. 2000).
We excluded all photon events that occurred during the bursts, reducing
effective on-target integration time from 16.8 ksec to 16.1 ksec.
Strong interstellar extinction and lack of co-alignment of the SiC
channels with the LiF channels prevented the collection of useful
data shortward of 1010 \AA .

We performed a simple ``collapse'' of the 2-dimensional spectral
image to a 1-dimensional spectrum, summing 100 pixels (LiF 1A
segment) or 120 pixels (LiF 2A) in each column, and we determined
the background level in a similar way.  This extraction compares
favorably to a subsequent pipeline processed spectrum.
In the regions of interest, we obtained a maximum S/N
of 20 per 15-pixel resolution element in the LiF 1A segment, and
30 in LiF 2A.  The wavelength solution was derived from narrow lines
in the diffuse cloud line of sight to LMC star Sk$-$67 111
(Shull et al.\ 2000).  Some residual scatter remained in the
wavelength solution ($\Delta v_{rms}$ $\approx$ 6 km s$^{-1}$),
precluding a detailed analysis of line velocities in this study.
Figure 1 depicts a portion of our spectrum, along with
our model of the H$_2$ lines described in the following paragraph.

The column densities for the $J$ = 0 and $J$ = 1 levels come from
direct fits to the profiles of the corresponding heavily damped lines.
The resolution of the {\it FUSE} spectra (R $\sim$ 12,000) is not
sufficient to separate the narrow, closely spaced components
contributing to the interstellar line profiles of the high-$J$ lines.
In view of the generally good correlation between the column densities
of \ion{Na}{1} and H$_2$ (Federman 1981; Welty \& Hobbs 2000) for N(Na I)
$\gtrsim$ 10$^{12}$ cm$^{-2}$,
we have used the five strongest \ion{Na}{1} components described in
\S 2 to model the H$_2$ lines.  The velocity structure for \ion{Na}{1}
was preserved for H$_2$, and the observed $b$-values for the
\ion{Na}{1} components were scaled to H$_2$ by the relation
\begin{displaymath}
b = \sqrt{\frac{2kT}{m} + v^2_{turb}}.
\end{displaymath}
Given the relatively narrow range of column densities, we have
assumed a one-to-one scaling between \ion{Na}{1} and H$_2$.  The
predicted H$_2$ component structure is given in Table 2.
From this structure we calculated a curve of growth and determined
the column densities for $J \geq$ 2 using 29 H$_2$
lines.  This curve of growth (similar to a single component
with $b$ = 3 km s$^{-1}$) is given in Figure 2, while
the column densities are given in Table 3.  The data in Tables 2
and 3 were used to calculate the model of the H$_2$ spectrum given
in Figure 1.

Lying on or near the saturation portion of the curve
of growth, the column densities for $J$ $\geq$ 4 are highly
sensitive to the assumed component structure.  If more than five
\ion{Na}{1} components trace significant amounts of H$_2$, the
larger ``effective'' $b$-value yields column densities for $J$ = 4
and $J$ = 5 decreased by an order of magnitude or more.  Similarly,
if the H$_2$ structure is less complex than we have assumed or if
components 4 and 5 trace less H$_2$ (as suggested by the CH data
of Gredel et al.\ 1993), the opposite situation occurs.
The $J$ = 5 lines would be shifted $\sim$0.5 dex to the damping
portion of the curve of growth, and the column densities
for the $J$ = 6 and $J$ = 7 lines could be increased by $\sim$0.5
dex or more.

\section{Discussion}

The total H$_2$ column density of 1.2 $\times$ 10$^{21}$ cm$^{-2}$
we have measured toward HD 73882 is statistically identical to the
largest previously observed via far-UV absorption spectroscopy,
1.1 $\times$ 10$^{21}$ cm$^{-2}$ toward HD 24534 (Mason, et al.\ 1976).
The molecular fraction $f$ = 2N(H$_2$)/[2N(H$_2$) + N(H I)] is $f$ = 0.65,
nearly identical to that found toward Zeta Oph ($f$ = 0.63; Bohlin,
et al.\ 1978).  The only known larger value, $f$ = 0.80, toward HD 24534
(Snow et al.\ 1998), will be redetermined via a planned {\it FUSE}
observation.

The abundances of CO, CN, CH, and C$_2$, relative to the total
hydrogen abundance, are more similar to those found in diffuse
clouds (i.e. toward Zeta Oph) than in dark clouds.  Because these
molecular abundances represent only a small fraction of the presumed
total carbon abundance, we conclude that the clouds toward HD 73882
have not reached the predicted transition point where carbon becomes
primarily molecular (e.g. van Dishoeck \& Black 1986).
The limited data presently available also suggest that the depletions
toward HD 73882 are similar to those found in cold diffuse clouds, an
issue we will explore in more detail in a future paper (Rachford et
al. 2000).

In principle, the CO abundance can be derived from our {\it FUSE}
spectrum.  However, the best candidate lines, from the C--X (0,0)
band at 1088 \AA, lie on the flat portion of the curve of growth
where the derived abundance is highly sensitive to the chosen
$b$-value.

Our derived H$_2$ column density, combined with the line-of-sight
extinction properties cited above, show that the ratio of hydrogen,
both molecular and total, to dust extinction, is also similar to the
value found for diffuse clouds (Bohlin et al.\ 1978).

The ratio of molecules in $J$ = 1 to $J$ = 0 (ortho-
to para-hydrogen) is usually interpreted as a measure of the gas
kinetic temperature, because in these clouds the collisional timescale
for depopulating these levels is shorter than the radiative timescale.
Our N(1)/N(0) ratio yields $T_{\rm kin} = 58 \pm 10$ K, consistent
with the average value $T_{\rm kin} = 77 \pm 17$ K found for diffuse
clouds with {\it Copernicus} data (Savage et al.\ 1977).  Shull
et al.\ (2000) find a somewhat larger value in a small sample of
{\it FUSE} data.

The excitation diagram in Figure 3 shows that the levels $J \geq 2$ in
this line of sight are not in thermal equilibrium at $T_{\rm kin}$,
but instead follow a different distribution characterized by an
excitation temperature, $T_{\rm ex} = 307 \pm 23$ K.  Non-thermal
excitation of the high-$J$ levels has been explained in terms of
UV pumping, in which the molecules cascade down through upper
rotational and vibrational lines following the absorption of a UV
photon (Black \& Delgarno 1973).  In this case, the excited H$_2$
may arise in the outer, optically thin regions of the cloud, rather
than in the self-shielded interior.  If so, the high-$J$
lines may exhibit small velocity shifts or enhanced $b$-values.
However, the limited spectral resolution and uncertain wavelength
scale of our {\it FUSE} data precludes a sensitive search for these
effects.

It is noteworthy that even in a line of sight thought to be
dominated by molecular cloud material within a single cloud complex,
the velocity structure is very complicated.  This structure, along
with the similarities between the current results and previous
results for diffuse clouds, suggests the possibility that we looking
through several Zeta Oph-type clouds rather than a single very
dense cloud.

Our analysis of the excited rotational levels of H$_2$
depends heavily on high-resolution ground-based optical absorption
and mm-wave emission line data to which we had access.  It is thus
imperative for future {\it FUSE} interstellar-line observations to
obtain appropriate high-resolution spectra in order to have any hope
of unambiguous interpretation of the lower resolution data.

\acknowledgments
We thank the referee, J. H. Black, for helpful comments.
This work is based on data obtained for the Guaranteed Time Team by the
NASA-CNES-CSA {\it FUSE} mission operated by the Johns Hopkins University.
Financial support to U.S. participants has been provided by
NASA contract NAS5-32985.

\clearpage

\begin{deluxetable}{lll}
\tablecaption{Ancillary Abundance Data for HD 73882}
\tablewidth{0pt}
\tablenum{1}
\tablehead{
\colhead{Species} & \colhead{N} & \colhead{Reference} \\
 & \colhead{(cm$^{-2}$)} &
}
\startdata
H I & 1.3 $\times$ 10$^{21}$ & Fitzpatrick \& Massa 1990 \\
K I & 2.4 $\times$ 10$^{11}$ & Welty et al.\ 2000 \\
Na I & 2.6 $\times$ 10$^{13}$ & Welty et al.\ 2000 \\
Ca II & 4.2 $\times$ 10$^{12}$ & Welty et al.\ 2000 \\
CH & 3.7 $\times$ 10$^{13}$ & Gredel et al.\ 1993 \\
CH$^+$ & 2.4 $\times$ 10$^{13}$ & Gredel et al.\ 1993 \\
C$_2$ & \phn\phd6 $\times$ 10$^{13}$ & Gredel et al.\ 1993 \\
CN & 3.8 $\times$ 10$^{13}$ & Gredel et al.\ 1993 \\
CO & 3.2 $\times$ 10$^{16}$ & van Dishoeck et al.\ 1991 \\
\enddata
\end{deluxetable}

\begin{deluxetable}{cccc}
\tablecaption{Assumed H$_2$ component structure}
\tablewidth{0pt}
\tablenum{2}
\tablehead{
\colhead{Comp.} & \colhead{Rel. strength} & \colhead{$b$} &
\colhead{$v_{\rm LSR}$} \\
& & \colhead{(km s$^{-1}$)} & \colhead{(km s$^{-1}$)}
}
\startdata
1 & 0.195 & 0.98 & +\phn2.62 \\
2 & 0.358 & 0.98 & +\phn4.20 \\
3 & 0.187 & 0.98 & +\phn5.83 \\
4 & 0.097 & 1.21 & +\phn7.56 \\
5 & 0.163 & 1.16 & +10.85 \\
\enddata
\end{deluxetable}

\begin{deluxetable}{ccc}
\tablecaption{Column density in each rotational state}
\tablewidth{0pt}
\tablenum{3}
\tablehead{
\colhead{$J$} & \colhead{log N$_J$} & \colhead{Error} \\
& \colhead{(cm$^{-2}$)} & \colhead{(cm$^{-2}$)}
}
\startdata
0 & 20.91 & 0.1 \\
1 & 20.59 & 0.1 \\
2 & 19.1\phn & 0.2 \\
3 & 18.5\phn & 0.3 \\
4 & 17.5\phn & 0.3\tablenotemark{a} \\
5 & 17.0\phn & 0.3\tablenotemark{a} \\
6 & 14.5\phn & 0.8\tablenotemark{a} \\
7 & 14.4\phn & 0.5\tablenotemark{a} \\
Total & 21.08 & 0.1 \\
\enddata
\tablenotetext{a}{The high-$J$ lines are subject to shifts larger
than the quoted errors with changes in the component structure;
see text.}
\end{deluxetable}

\clearpage

\begin{figure}
\epsscale{0.65}
\plotone{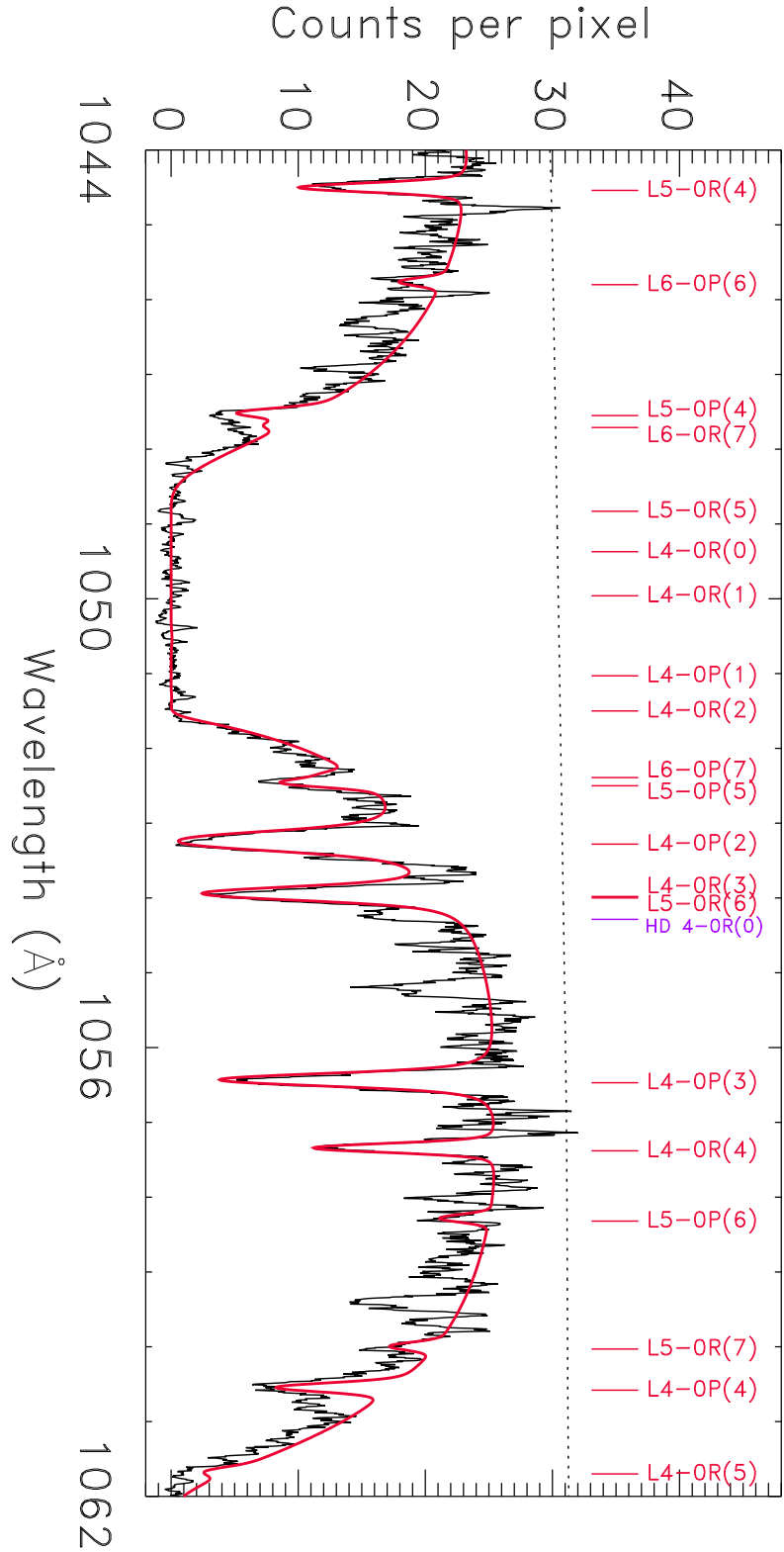}
%\plotfiddle{spec.eps}
\caption{Portion of LiF 1A spectrum with H$_{2}$ model overlaid.  The
dotted line depicts the continuum derived from modeling the overlapping
wings of the lines from different bands.  The HD line is not
included in the model (see Ferlet, et al.\ 2000 for the HD analysis).}
\end{figure}

\begin{figure}
\plotone{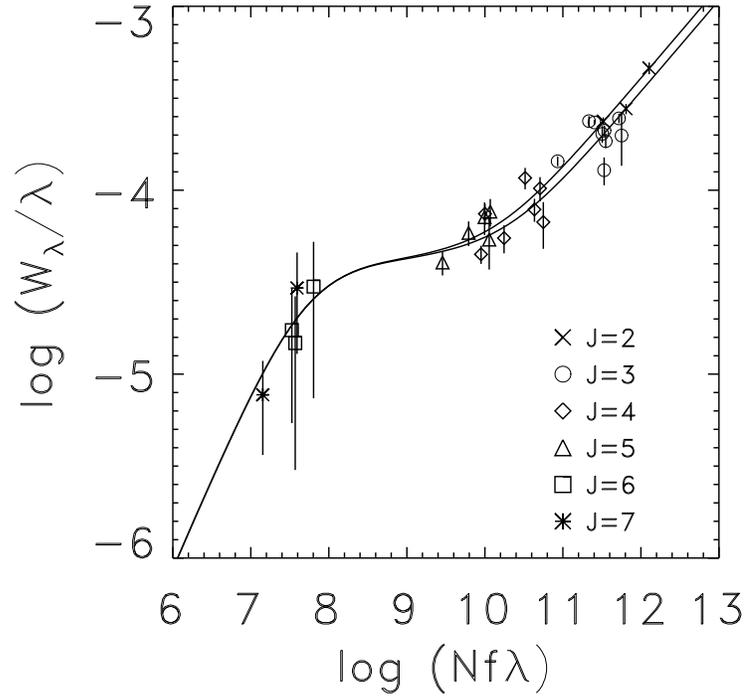}
\caption{Multi-component curve of growth for HD 73882.  The two
curves represent the range in damping constants for the measured
lines.  The error bars lines only
give the formal uncertainty in the line fits and may underestimate
the total uncertainties for the stronger lines.}
\end{figure}

\begin{figure}
\plotone{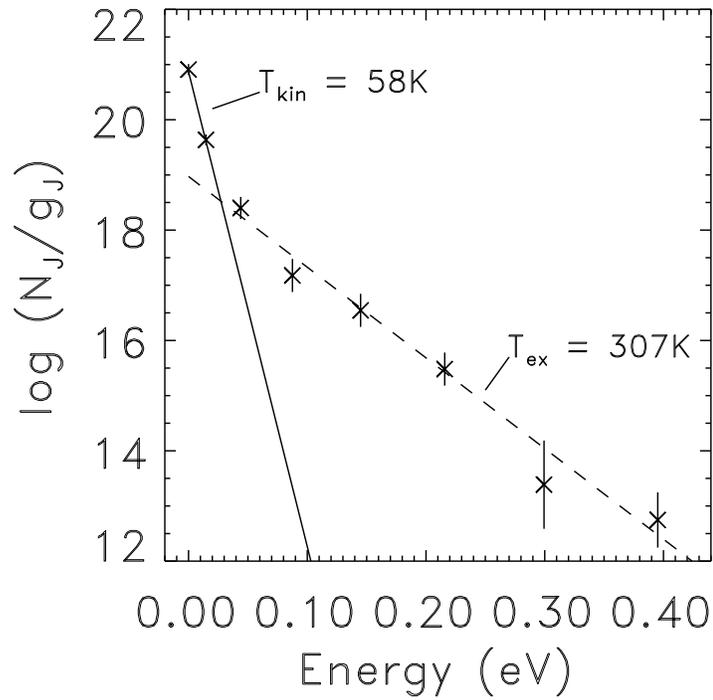}
\caption{Excitation plot for HD 73882 for $J$ = 0
through $J$ = 7.  The solid line corresponds to the assumed kinetic
temperature derived from $J$ = 0 and $J$ = 1, while the
dashed line gives a weighted fit to the $J$ $\geq$ 2 lines.}
\end{figure}

\end{document}